\documentclass[a4paper,fleqn,usenatbib]{mnras}

\usepackage{graphicx}	
\usepackage{amsmath}	
\usepackage{amssymb}	
\usepackage{savesym}
\usepackage{natbib}

\usepackage{mathrsfs,dsfont}
\usepackage{multirow}
\usepackage{captcont,subcaption}
\usepackage{float}
\usepackage{booktabs}
\usepackage{epstopdf}
\usepackage{color}
\usepackage{hyperref}
\usepackage{verbatim}

\usepackage[none]{hyphenat}

\graphicspath{{figures/}}

\def\be{\begin{equation}}
\def\ee{\end{equation}}
\def\kms{{\rm \,km\,s^{-1}}}

\def\Gyr{{\rm \,Gyr}}

\def\kpc{{\rm \,kpc}}

\def\keV{{\rm \,keV}}

\def\msun{{\,M_\odot}}

\title[Standoff Distance of Bow Shocks in Galaxy Clusters]{Standoff Distance of Bow Shocks in Galaxy Clusters as Proxy for Mach Number}

\author[Congyao Zhang et al.]{
Congyao Zhang,$^1$\thanks{E-mail: cyzhang@mpa-garching.mpg.de}
Eugene Churazov,$^{1,2}$
William R. Forman,$^{3}$
Christine Jones$^{3}$
\\
$^1$~Max Planck Institute for Astrophysics, Karl-Schwarzschild-Str. 1, D-85741 Garching, Germany  \\
$^2$~Space Research Institute (IKI), Profsoyuznaya 84/32, Moscow 117997, Russia \\
$^3$~Smithsonian Astrophysical Observatory, Harvard-Smithsonian Center for Astrophysics, 60 Garden St., Cambridge, MA 02138 \\
}

\date{Accepted XXX. Received YYY; in original form ZZZ}

\pubyear{2018}

\begin{document}
\label{firstpage}
\pagerange{\pageref{firstpage}--\pageref{lastpage}}
\maketitle

\begin{abstract}
{X-ray observations of merging clusters provide many examples of bow shocks leading merging subclusters. While the Mach number of a shock can be estimated from the observed density jump using Rankine-Hugoniot condition, it reflects only the velocity of the shock itself and is generally not equal to the velocity of the infalling subcluster dark matter halo or to the velocity of the contact discontinuity separating gaseous atmospheres of the two subclusters. Here we systematically analyze additional information that can be obtained by measuring the standoff distance, i.e. the distance between the leading edge of the shock and the contact discontinuity that drives this shock. The standoff distance is influenced by a number of additional effects, e.g. (1) the gravitational pull of the main cluster (causing acceleration/deceleration of the infalling subcluster), (2) the density and pressure gradients of the atmosphere in the main cluster, (3) the non-spherical shape of the subcluster, and (4) projection effects. The first two effects tend to bias the standoff distance in the same direction, pushing the bow shock closer to (farther away from) the subcluster during the pre- (post-)merger stages. Particularly, in the post-merger stage, the shock could be much farther away from the subcluster than predicted by a model of a body moving at a constant speed in a uniform medium. This implies that a combination of the standoff distance with measurements of the Mach number from density/temperature jumps can provide important information on the merger, e.g. differentiating between the pre- and post-merger stages.}

\end{abstract}

\begin{keywords}
hydrodynamics -- shock waves -- methods: numerical -- galaxies: clusters: intracluster medium -- X-rays: galaxies: clusters
\end{keywords}


\section{Introduction} \label{sec:introduction}

\begin{figure}
\centering
    \begin{subfigure}[t]{0.45\textwidth}
        \centering
        \includegraphics[width=\linewidth]{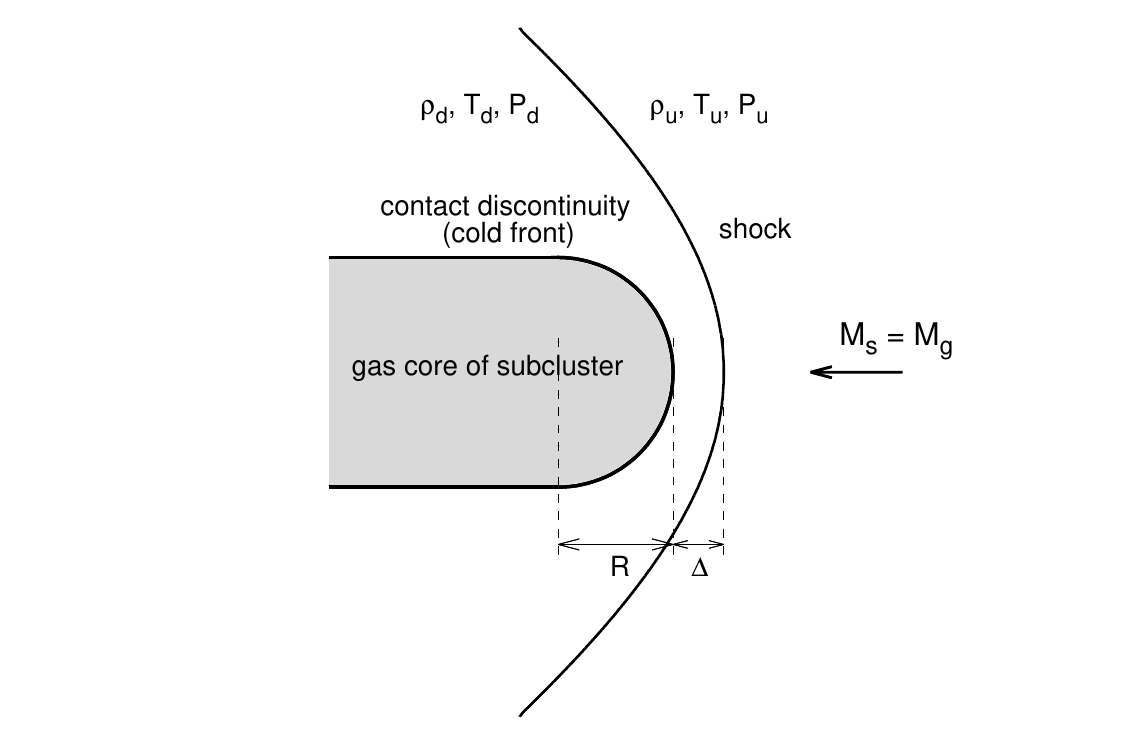}
        \caption{}
        \label{fig:scheme_merger}
    \end{subfigure}
    \begin{subfigure}[t]{0.45\textwidth}
        \centering
        \includegraphics[width=\linewidth]{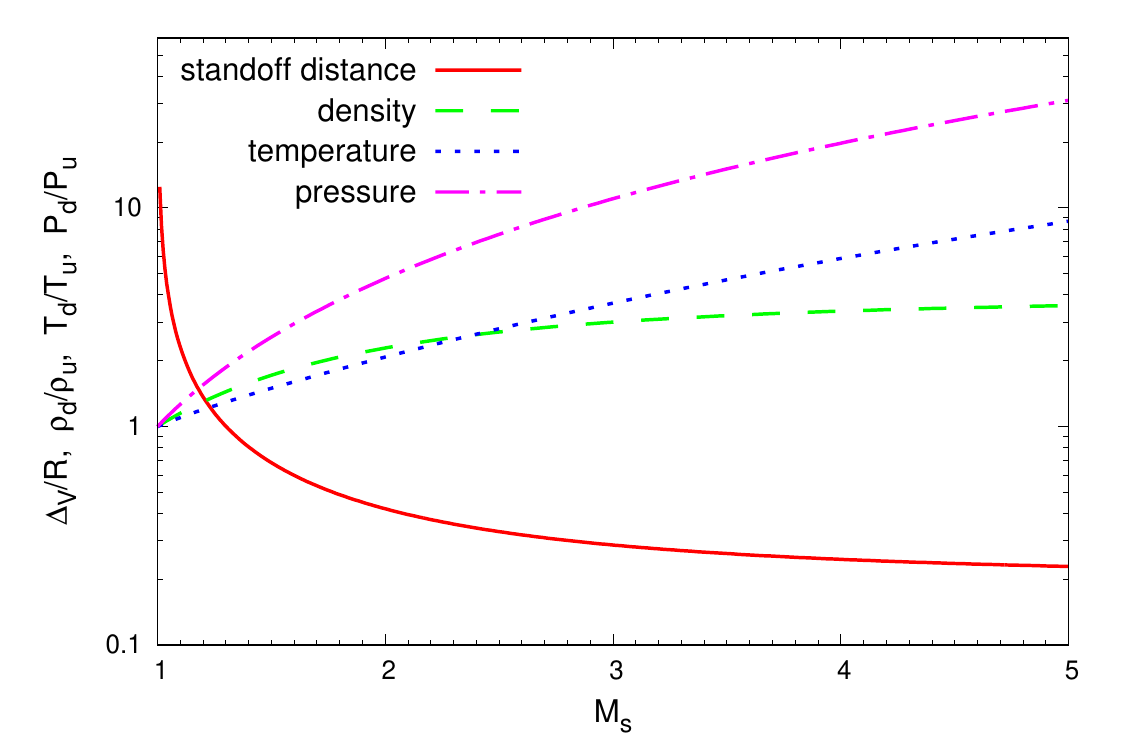}
        \caption{}
        \label{fig:mach_proxy}
    \end{subfigure}
\caption{\textit{Panel (a):} sketch showing a bow shock driven by the infalling subcluster (in the rest frame of the shock). The cold gas core of the subcluster is enveloped by a sharp contact discontinuity (a.k.a. cold front). The standoff distance $\Delta$ and the curvature radius of the cold front $R$ are marked in the figure. \textit{Panel (b):} various observational proxies for the shock Mach number (see Appendix~\ref{sec:appendix:proxy}). The three curves, increasing with Mach number, correspond to the RH relations between the gas properties (density, temperature, and pressure) on the downstream and upstream sides of the shock. The decreasing curve is the normalized standoff distance $\Delta_V/R$ from \citet{Verigin2003} for a moving sphere (see equation~\ref{eq:proxy_delta}). Unlike the RH relations, the standoff distance is an unambiguous function of $\mathcal{M}_s$, only if the sphere is moving at a constant velocity through a homogeneous medium, i.e. $\mathcal{M}_g=\mathcal{M}_s$.
}
%
\end{figure}

In the hierarchical structure formation scenario, the growth of galaxy clusters occurs via mergers of less massive structures (e.g. galaxy groups, clusters). Shock waves are naturally formed in this process since the infalling gas haloes usually move faster than the local speed of sound ($c_s$, see e.g. \citealt{Markevitch2007,Bykov2015} for reviews). High-angular-resolution X-ray images are able to identify sharp density and temperature discontinuities of these shocks in the intracluster medium (ICM). Dozens of shocks (or shock candidates) have been discovered so far in this way (e.g. \citealt{Markevitch2002,Markevitch2005,Botteon2018}). The shock Mach number ($\mathcal{M}_s$, typically $\mathcal{M}_s\lesssim4$) is subsequently determined by applying the Rankine-Hugoniot (RH) conditions to the derived gas density and temperature jumps at the shock front (see Fig.~\ref{fig:mach_proxy}). Under the assumption of a steady motion in a homogeneous medium, the shock Mach number coincides with the Mach number of the infalling gas halo ($\mathcal{M}_g$) relative to the atmosphere of the main cluster, i.e. $\mathcal{M}_s=\mathcal{M}_g$. As we discuss below, this equality of the Mach numbers is not necessarily true in the merging clusters. Instead, the shock, the infalling dark matter (DM) halo, and the gas of the subhalo can have appreciably different velocities. Therefore, more parameters are needed to characterize the merger in a more comprehensive way.

The geometry of the shock (e.g. Mach angle, standoff distance), may provide independent ways of measuring the Mach number \citep{Vikhlinin2001,Markevitch2002,Russell2010,Wezgowiec2011,Hallman2018}. In these studies, the infalling gas core (colder and denser than the gas in the main cluster) is usually taken as a solid body that diverts the flow and forms a bow shock in front (see a sketch in Fig.~\ref{fig:scheme_merger}). The theory of bow shocks driven by blunt bodies has been extensively studied theoretically and experimentally, including space physics applications \citep[e.g.][]{Dyke1958,Farris1994,Fairfield2001,Petrinec2002,Verigin2003,Keshet2016}.
The standoff distance $\Delta$ is defined as the distance between the stagnation point of the body and the closest point on the shock front (see Fig.~\ref{fig:scheme_merger}). In galaxy clusters, it is usually measured as the distance between the cold and shock fronts (see Fig.~\ref{fig:scheme_merger}). In this work, we use the results of \citet{Verigin2003} as a baseline model for the standoff distance (hereafter, $\Delta_V$) for a moving sphere with a radius $R$ (see equation~\ref{eq:proxy_delta}). The normalized standoff distance $\Delta_{V}/R$ is a monotonically decreasing function of the Mach number $\mathcal{M}_s=\mathcal{M}_g$ (particularly sensitive to $\mathcal{M}_s$ when $\mathcal{M}_s\lesssim3$, see Fig.~\ref{fig:mach_proxy}).

For galaxy clusters, measuring $\Delta$ from X-ray data is straightforward. The normalized standoff distance $\Delta/R$ is a pure geometrical quantity and there is no need to resolve the structure of the entire Mach cone (here, $R$ refers to the curvature radius of the cold front). It is, therefore, an attractive geometrical proxy for the shock Mach number. However, it has its own issues when applied to real merging clusters. For instance, significant discrepancies between the Mach numbers measured from the standoff distance and the jump conditions are found in many clusters \citep[][and references therein]{Dasadia2016}, which typically show much larger standoff distances than the theoretical expectations, given the shock Mach number (e.g. derived from the X-ray surface brightness jump). These discrepancies might be caused by a variety of reasons, including but not limited to:
\begin{itemize}
  \item Continuous acceleration or deceleration of the subcluster when moving in the potential well of the main halo, i.e. the motion is not stationary.
  \item Motion of the subcluster through the gas with substantial density (pressure) gradients along the direction of motion.
  \item Non-trivial (and time variable) shape of the infalling subcluster and projection effects that lead to ambiguities in determining the standoff distance and the curvature radius of the cold front.
\end{itemize}

Therefore, whether the standoff distance could be a robust proxy of shock Mach number is an open question. At the same time, the discrepancy between the shock Mach number derived from the Rankine-Hugoniot condition and the Mach number derived from the standoff distance has an important diagnostic power that can be used to better characterize the merger configuration. The aim of this work is to systematically investigate various effects that influence the estimates of the Mach number from the standoff distance in galaxy clusters. To isolate issues related to the evolving shape of the subcluster, we mainly concentrate on a model of a rigid body moving in a static potential well.

This paper is organised as follows. In Section~\ref{sec:sph}, we use smoothed-particle hydrodynamics (SPH) simulation to illustrate the evolution of the standoff distance during the merger process. We demonstrate that the rigid body approximation captures many, but admittedly not all, important issues associated with using the standoff distance as a proxy for the Mach number. Our main results are presented in Section~\ref{sec:rigid}, where we explore the impact of the gravitational potential, subcluster shape, and projection effects on the standoff distance, respectively. In Section~\ref{sec:conclusion}, we summarize our findings.

\section{Illustrative SPH simulation of a minor merger} \label{sec:sph}

Before moving to a rigid-body  model, we first use an SPH simulation to illustrate the (minor) merger\footnote{We have also tested a case of major merger with the mass ratio $\xi=2$, which generally shows very similar results with those of the minor merger.} process between two idealized galaxy clusters. Each cluster consists of a spherical dark matter halo and gas halo (see Appendix~\ref{sec:appendix:sph} for the detailed description of the simulation method). This simulation is aimed to provide clues to the standoff distance issues found in observations.

\begin{figure*}
\centering
\includegraphics[width=0.9\textwidth]{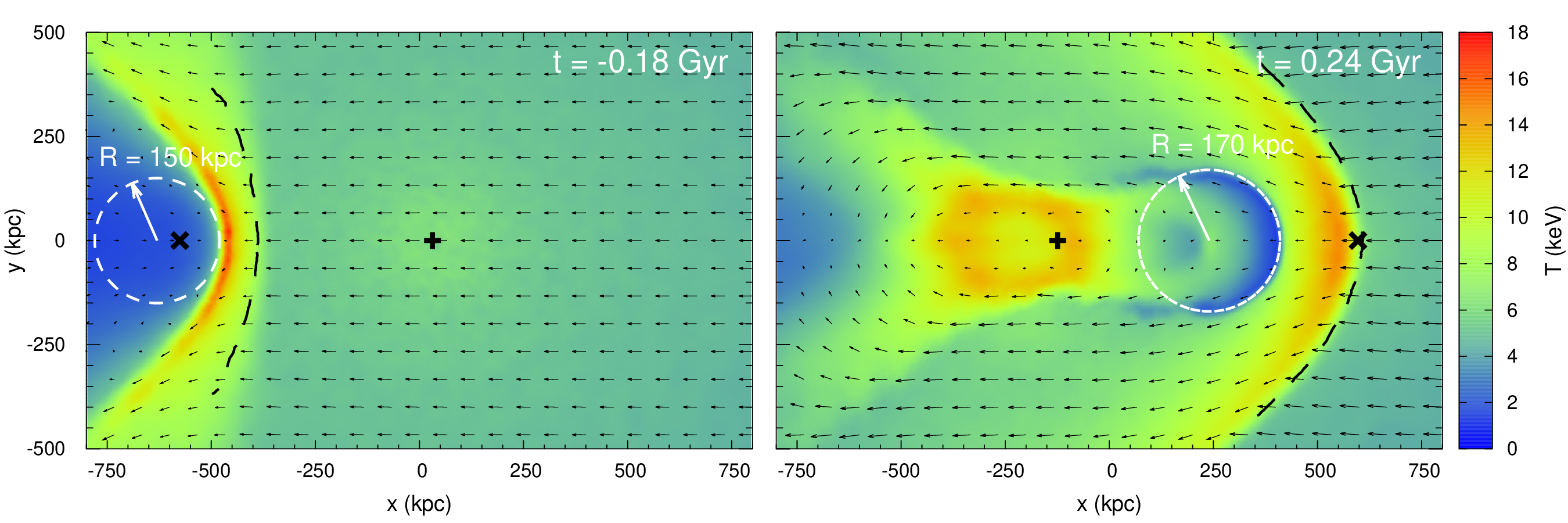}
\caption{
Temperature slices of a (minor) merger of two idealized galaxy clusters at the moment before (left panel; $t=-0.18\Gyr$) and after (right panel; t=0.24\Gyr) the first pericentric passage (defined as $t=0$). The overplotted arrows show the gas velocity field in the rest frame of the shock. The black `$+$' and `$\times$' symbols mark the positions of the main cluster's and subcluster's centers (viz. peak of the mass density), respectively. The white circles with radii $R=150$ and $170\kpc$ show the adopted approximation to the size of the cold gas core in the infalling subcluster. The black dashed curves illustrate the shape and position of the shock fronts, whose Mach numbers are $1.9$ and $2.4$ in the left and right panels, respectively. Even though the shock Mach number is larger in the right panel, its shock is farther away from the cold front than in the left panel. This clearly illustrates the difference in the standoff distances before and after the pericentric passage (see Section~\ref{sec:sph}). }
\label{fig:sph_slice}
\end{figure*}

\begin{figure*}
\centering
    \begin{subfigure}[t]{0.45\textwidth}
        \centering
        \includegraphics[width=\linewidth]{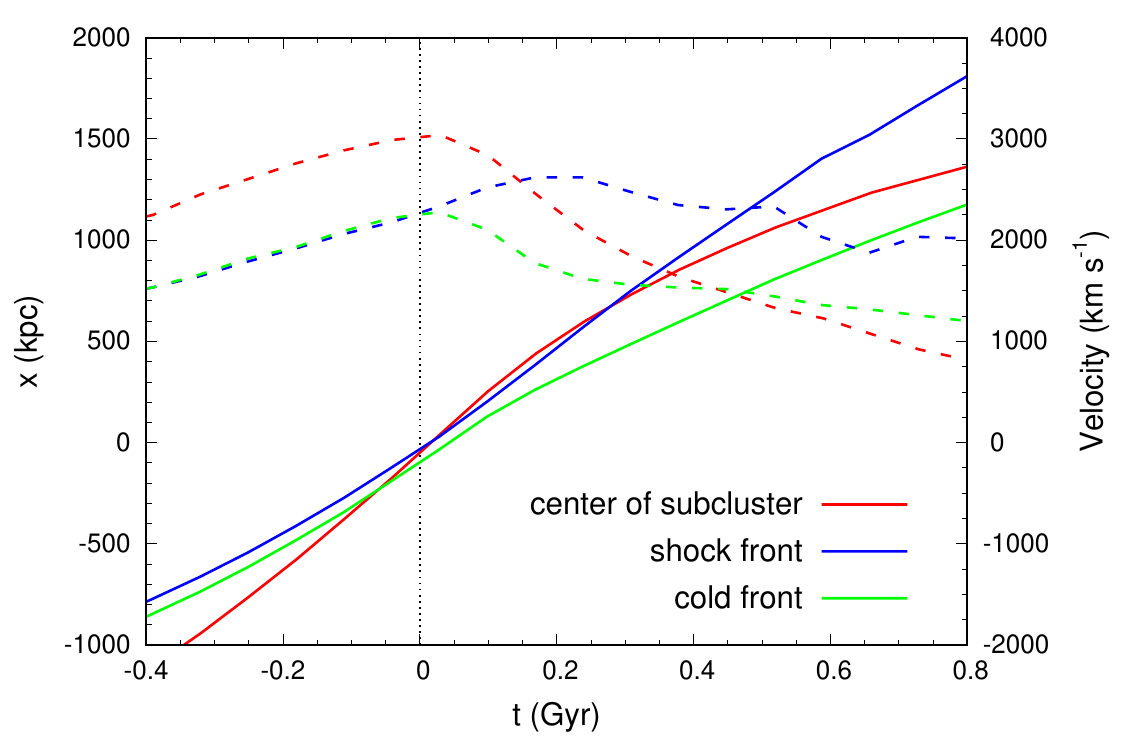}
    \end{subfigure}
    \begin{subfigure}[t]{0.45\textwidth}
        \centering
        \includegraphics[width=\linewidth]{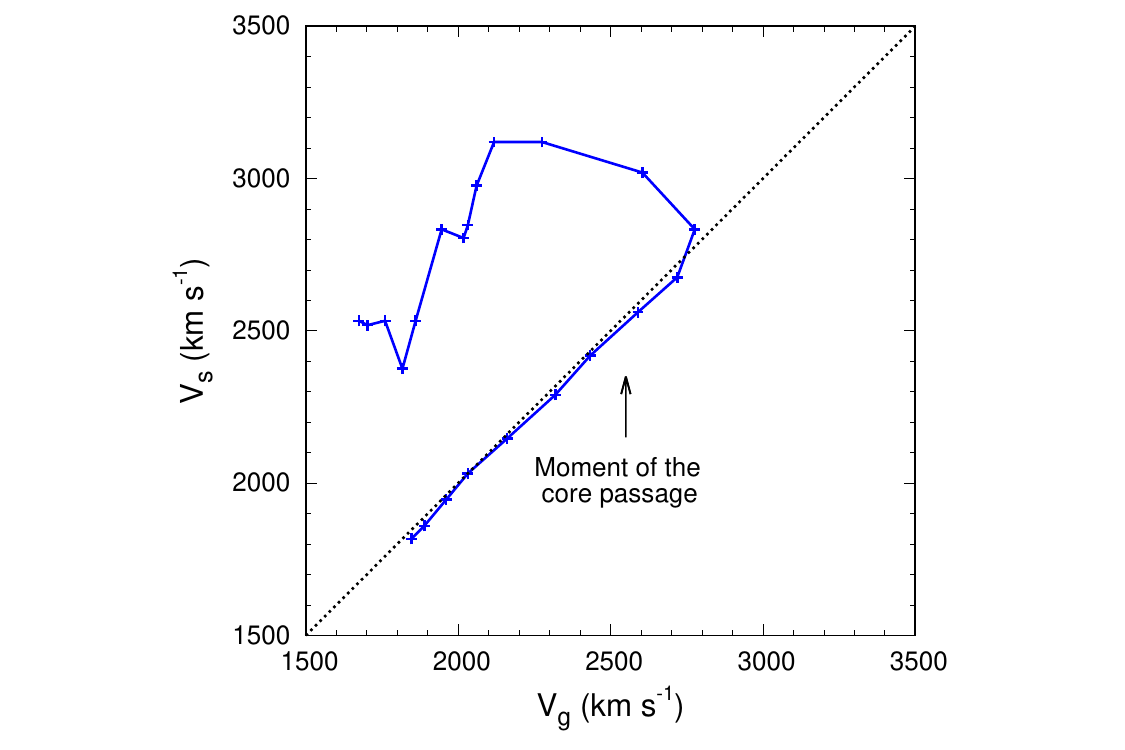}
    \end{subfigure}
\caption{\textit{Left panel:} time evolution of the positions (solid lines) and velocities (dashed lines) along the merger axis of the subcluster (peak of the mass surface density), shock front, and cold front, from the SPH simulation. All positions and velocities are relative to the center of mass of the merging cluster. The difference in velocities of the subcluster (dashed red) and the cold front (dashed green) before the pericentric passage ($t=0$, marked by a vertical dotted line) is due to the ram pressure that pushes the gas in the subcluster in the direction opposite to the velocity of the subcluster. \textit{Right panel:} the shock velocity $V_s\,(\equiv\mathcal{M}_sc_s)$ as a function of the velocity of the cold front $V_g$ (relative to the unshocked gas in front of the shock). Before the core passage ($t<0$), the velocities of the subcluster and the bow shock are close to each other; while after the core passage, the subcluster `lags' behind the shock. This is caused by two effects: (1) gravitational attraction of the main cluster that decelerates the subcluster after core passage; (2) the negative density gradient of the atmosphere that tends to make the shock faster and stronger (see Section~\ref{sec:sph}). }
\label{fig:sph_evolution}
\end{figure*}

Figure~\ref{fig:sph_slice} shows the temperature slices (taken in the merger plane) of our simulation (a minor merger with the mass ratio $\xi=10$, see Table~\ref{tab:sph_params} for merger parameters). The left and right panels show the moments before and after the primary pericentric passage of the clusters (i.e. pre- and post-merger stages)\footnote{For clarity, we set the evolution time $t = 0$ at the primary pericentric passage.}, respectively, where both the contact discontinuity (cold front) and the shock (black dashed curves)\footnote{In Figures~\ref{fig:sph_slice} and \ref{fig:sph_evolution}, locations of the cold and shock fronts are determined as the local steepest positions in the gas temperature profile along the merger axis.} are clearly seen. The overplotted arrows illustrate the gas velocity field in the rest frame of shock.

In the {\it pre-merger stage} (left panel in Fig.~\ref{fig:sph_slice}), the velocity difference between the shock and the cold gas core is small (see also the blue and green dashed lines in the left panel of Fig.~\ref{fig:sph_evolution}), i.e. the shock is moving with the velocity of the gas halo of the subcluster ($\mathcal{M}_s\simeq\mathcal{M}_g$). The normalized standoff distance $\Delta/R\sim0.5$ agrees with the expectation for a moving sphere shown in Fig.~\ref{fig:mach_proxy} for the shock Mach number estimated from the temperature discontinuity, i.e. $\mathcal{M}_{s}\simeq1.9$. Here we assume $R=150\kpc$ (see the white circle in the left panel of Fig.~\ref{fig:sph_slice}). We note in passing that, the shape of the subcluster is obviously non-spherical (see also Fig.~\ref{fig:sph_shape} below). The impact of the subcluster shape on the standoff distance will be discussed in Section~\ref{sec:rigid:shape}.

In the {\it post-merger stage} (right panel in Fig.~\ref{fig:sph_slice}), however, the relative velocity between the shock and the infalling gas core is large, up to $\sim800\,\kms$ (see Fig.~\ref{fig:sph_evolution}). The normalized standoff distance estimated from Figure~\ref{fig:sph_slice} ($\Delta/R\sim 1.2$, where $\Delta\simeq200\kpc$ and $R\simeq170\kpc$ at $t=0.24\Gyr$) is significantly larger than the expected value $\Delta_V/R\sim0.34$ for $\mathcal{M}_{s}\simeq2.4$ (see Fig.~\ref{fig:mach_proxy}). \citet{Springel2007} reported a similar phenomenon in their models of the `Bullet' cluster.

Figure~\ref{fig:sph_evolution} shows more clearly the positions and velocities (along the merger axis) of the subcluster (peak of the mass surface density), the contact discontinuity (cold front) and the shock (see the left panel). Here, all quantities are measured relative to the center of mass of the merging cluster. In particular, the shock velocity ($\mathcal{M}_sc_s$) as a function of the velocity of the cold front (relative to the gas in front of the shock) is shown in the right panel. We note that the motion of the subcluster center reveals a different behavior from that of the cold front/shock, due to the collisionless nature of the DM. From these plots, it is clear that the velocities of the cold front and the shock begin to deviate from each other after the core passage. As we argue below, the apparent lag of the subcluster relative to the shock is predominantly due to (1) gravitational attraction of the main cluster that decelerates the subcluster after core passage; (2) the negative density gradient of the atmosphere that tends to make the shock stronger \citep[e.g.][]{Sarazin2016}. We will explore the impact of these effects on the standoff distance in the next section.

The SPH simulations shown in  Figure~\ref{fig:sph_slice} further motivated us to treat the infalling gas halo as a rigid blunt body.  While the Kelvin-Helmholz instability could be severely suppressed in SPH simulations, it mainly occurs near the sides and/or in the wake of the gas core (\citealt{Zuhone2016}; see more discussions on the ram-pressure stripping in \citealp{Roediger2015}), so the leading edge of the cool core is expected to maintain a relatively smooth shape. This shape does change with time, but relatively slowly (see also Fig.~\ref{fig:sph_shape} below). The rigid-subcluster model is therefore expected to be a fair approximation for the infalling gas core for the purpose of studying the standoff distance.

\section{Rigid-subcluster Model} \label{sec:rigid}

\begin{figure}
\centering
\includegraphics[width=0.45\textwidth]{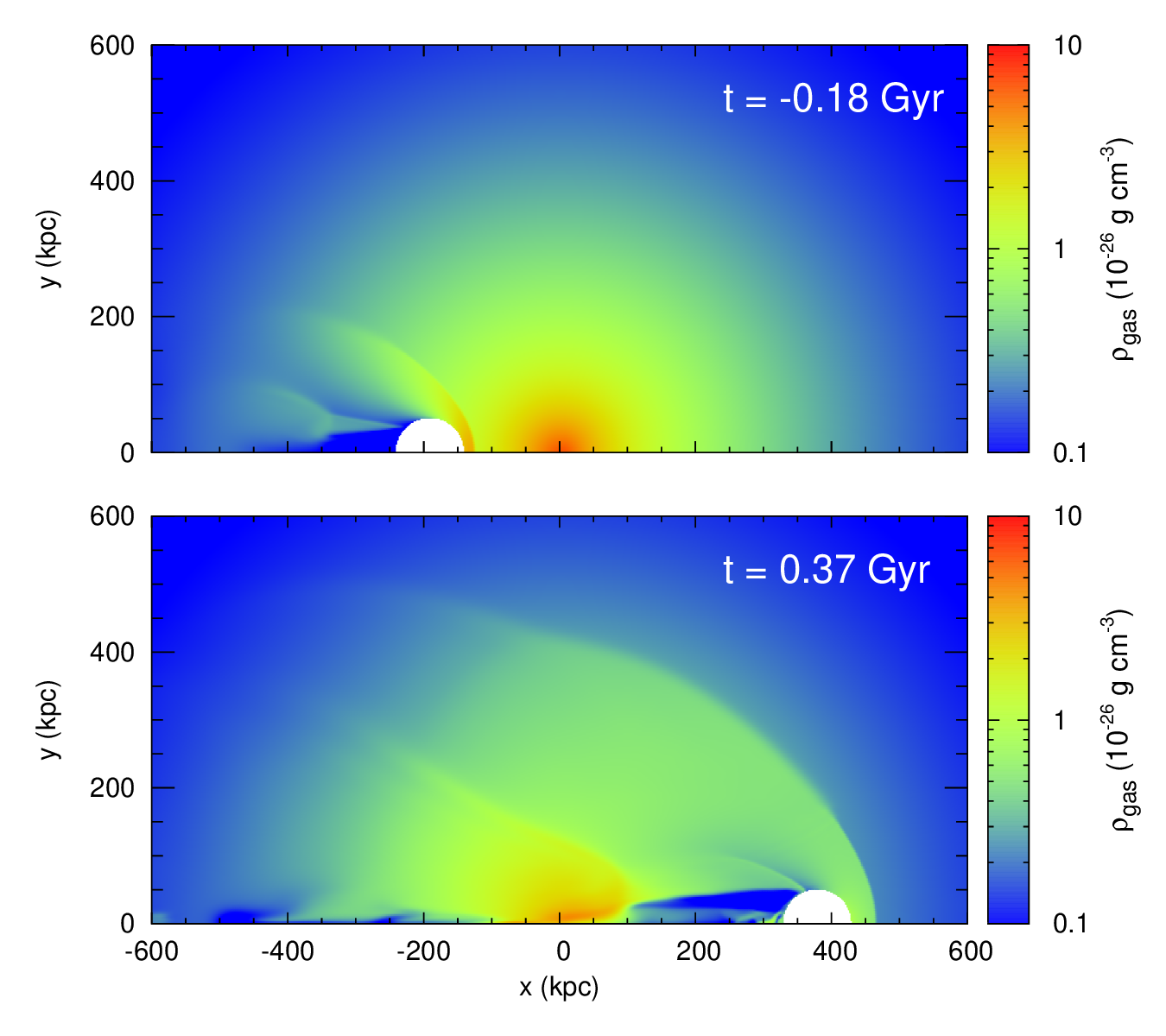}
\caption{Snapshots of the gas density of the G2 run (a sphere moving with constant velocity through the isothermal atmosphere with varying density and pressure). This figure shows an example of a rigid-subcluster simulation. The shape and position of the shock front are modified by the density and pressure gradients (see Section~\ref{sec:rigid}).}
\label{fig:foam_slice}
\end{figure}

In this section, we consider a `rigid-subcluster' model, which is expected to capture some of the effects associated with the bow shock driven by the infalling subcluster. Essentially, we are studying the bow shocks driven by a rigid body, but allow for variations of the body velocity and/or ambient gas density and pressure, while the impact of the shape evolution is ignored. Since the shape and motion of the rigid subclusters are highly controllable in this model, we are able to single out the influence of each parameter and compare the results to our reference model of a sphere steadily moving in the homogeneous medium. Numerical aspects of the rigid-subcluster model are described in Appendix~\ref{sec:appendix:openfoam} (see Table~\ref{tab:foam_params} for the simulation parameters). An example of these simulations is shown in Figure~\ref{fig:foam_slice}. In addition, projection effects, which are important for the analysis of observations, are also considered in this section. 

\subsection{Role of Gravitational Potential} \label{sec:rigid:gravity}

The gravitational potential of the main cluster plays a key role in affecting the standoff distance of the shock driven by the infalling subcluster. On the one hand, it accelerates (or decelerates) the infalling subcluster and causes a relative motion between the shock and cold fronts (see Fig.~\ref{fig:sph_evolution}); on the other hand, it affects the strength and the propagation velocity of the shock itself via the spatially variable density and pressure profiles. We therefore perform two groups of simulations to explore both of these effects (see Table~\ref{tab:foam_params}), viz. (1) V-group: the rigid subcluster moves in a uniform medium with time-dependent velocity; (2) G-group: the rigid subcluster moves with constant (or time-dependent) velocity in a static gravitational potential well holding an isothermal atmosphere. We emphasize that, in the G-group simulations, the velocity of the moving subcluster is set by hand and the role of the gravitational potential is solely to keep the atmosphere in equilibrium.

\subsubsection{Moving along and opposite to the density and pressure gradients} \label{sec:rigid:gravity:gradient}

\begin{figure}
\centering
\includegraphics[width=0.45\textwidth]{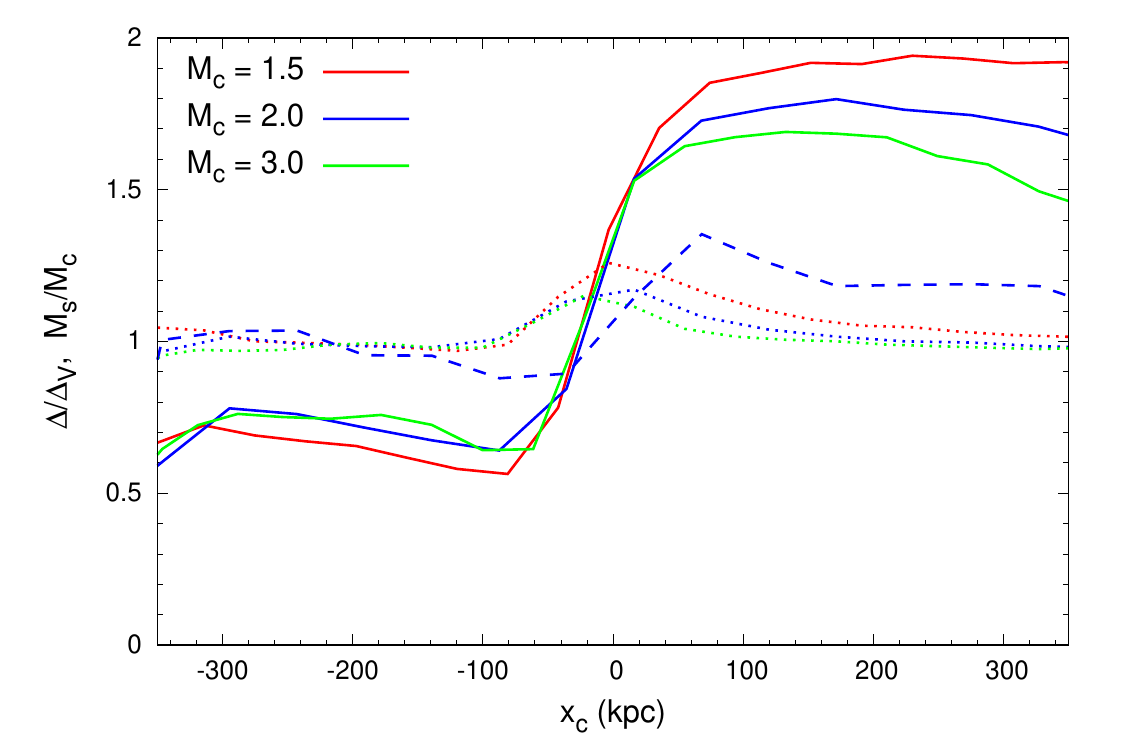}
\caption{Impact of spatially varying density (pressure) profiles on the standoff distance for a sphere moving at a constant velocity. The curves show the ratio between the standoff distance obtained in simulations and the baseline value $\Delta/\Delta_{V}$ in the G1/G2/G3 (solid lines, $R=50\kpc$) and the G4 (dashed line, $R=12.5\kpc$) runs. The dotted lines show the ratio between the shock and subcluster Mach numbers $\mathcal{M}_s/\mathcal{M}_c$ in the G1/G2/G3 runs. The horizontal axis represents the horizontal position of the subcluster center $x_{\rm c}$ (similarly in Figs.~\ref{fig:foam_dist_uni} and \ref{fig:foam_mach_time} below). The reason for significant deviations of $\Delta$ from the reference value is clear: for $x_{\rm c}\lesssim 0$, an increasing upstream gas density/pressure (compare to the unshocked gas on the downstream side) tends to slow down the shock and push the front closer to the body (a case of a subcluster falling towards the potential well of the main cluster). For the body moving in the direction of decreasing density and pressure  ($x_{\rm c}\gtrsim0$), this trend is obviously reversed. The G4 run (dashed line) shows that if the size of the body is small compared to the scale length of the atmosphere gradient, then $\Delta$ approaches the baseline value $\Delta_{V}$ (see Section~\ref{sec:rigid:gravity:gradient}).}
\label{fig:foam_dist_icm}
\end{figure}

Figure~\ref{fig:foam_dist_icm} shows the evolution of the ratio $\Delta/\Delta_{V}$ from the G-group simulations (solid and dashed lines; $\mathcal{M}_s$ is used in equation~\ref{eq:proxy_delta} to evaluate $\Delta_{V}$). The density gradient in the atmosphere leads to a deviation of the standoff distance from the  reference value. We find $\Delta/\Delta_{V}<1$ if the subcluster moves in the direction of the density gradient and $\Delta/\Delta_{V}>1$ otherwise. The deviations, up to $50\%$ in G1/G2/G3 runs, are nearly independent on the Mach number of the rigid subcluster ($\mathcal{M}_c$)\footnote{In the SPH simulation, the Mach number of the infalling gas core is different from that of the subcluster (see Fig.~\ref{fig:sph_evolution}). In the rigid-subcluster simulations, however, both Mach numbers have a same value, i.e. $\mathcal{M}_g=\mathcal{M}_c$.}. The reason for these deviations is clear: an increasing upstream gas density and pressure (in comparison to the unshocked gas on the downstream side) tend to slow the shock and push the front closer to the body (a case of a subcluster falling towards the potential well of the main cluster). For the body moving in the direction of decreasing density and pressure, this trend is obviously reversed. The evolution of the ratio between the shock and subcluster Mach numbers $\mathcal{M}_s/\mathcal{M}_c$ is shown in Figure~\ref{fig:foam_dist_icm} (dotted lines), consistent with the above arguments. Since the difference in density and pressure increases with the size of the body, it is reasonable to characterize this effect by the ratio between the pressure scale height of the atmosphere $H_{\rm P}\,(\equiv|{\rm d}\ln P_{\rm gas}/{\rm d}r|^{-1})$ and the subcluster size $R$, where $P_{\rm gas}$ is the gas pressure. In our simulations, $H_{\rm P}$ varies between $\sim50$ and $\sim250\kpc$ from the cluster center to the outskirts (i.e. $x=400\kpc$; see equation~\ref{eq:profile_rho}). The blue solid and dashed lines in Figure~\ref{fig:foam_dist_icm} compare $\Delta/\Delta_{V}$ for subclusters moving at $\mathcal{M}_c=2$ but with different sizes $R=50$ and $12.5\kpc$ (G2 and G4 runs). One can see that density (pressure) gradients play a significant role when $H_{\rm P}/R$ is less than a factor of a few.

\subsubsection{Acceleration or deceleration of the subcluster} \label{sec:rigid:gravity:acceleration}

Figure~\ref{fig:foam_dist_uni} shows the variations of the normalized standoff distance in the V-group runs (solid lines), where the subcluster is moving with variable velocity through the uniform medium. Here we keep a constant amplitude $a_{\rm x}$ for the subcluster acceleration $a_{\rm c}(t)$ but its direction inverts at the moment $t_{\rm p}$ when the subcluster center reaches $x_{\rm c} = 0$, i.e.
\be
{a}_{\rm c}(t) = \left\{
  \begin{array}{lr}
    ~a_{\rm x},     & t \leq t_{\rm p}\\
   -a_{\rm x},   & t > t_{\rm p}
  \end{array}
\right.,
\label{eq:profile_vel}
\ee
see Table~\ref{tab:profile_vel} for the adopted parameters. The corresponding velocity profiles are shown in the inset. For comparison, the corresponding normalized standoff distances from the baseline model are shown with the dotted lines. The differences between the simulation results and the baseline model illustrate the non-steady character of the bow shocks formed in front of the accelerating/decelerating body. The shocks need time to respond to the change of the obstacle's velocity. This effect becomes very prominent as the Mach number of the body approaches unity.
Even when the subcluster decelerates to a subsonic velocity, the shock still survives and propagates away from the subcluster (i.e. from the cold front, see the right half of Fig.~\ref{fig:foam_dist_uni}). It also implies that the shock Mach number may not reflect the velocity of the subcluster accurately. Thus, for a subcluster still falling into the potential well, the continuous acceleration will tend to decease $\Delta$, while after the core passage this trend is reversed. We finally stress that our quantitative results depend on the initial position of the rigid subcluster (set at $x_{\rm c}=-450\kpc$ for all our rigid-subcluster simulations, see Appendix~\ref{sec:appendix:openfoam}).

\subsubsection{Acceleration/deceleration plus density (pressure) gradients} \label{sec:rigid:gravity:combination}

\begin{figure}
\centering
\includegraphics[width=0.45\textwidth]{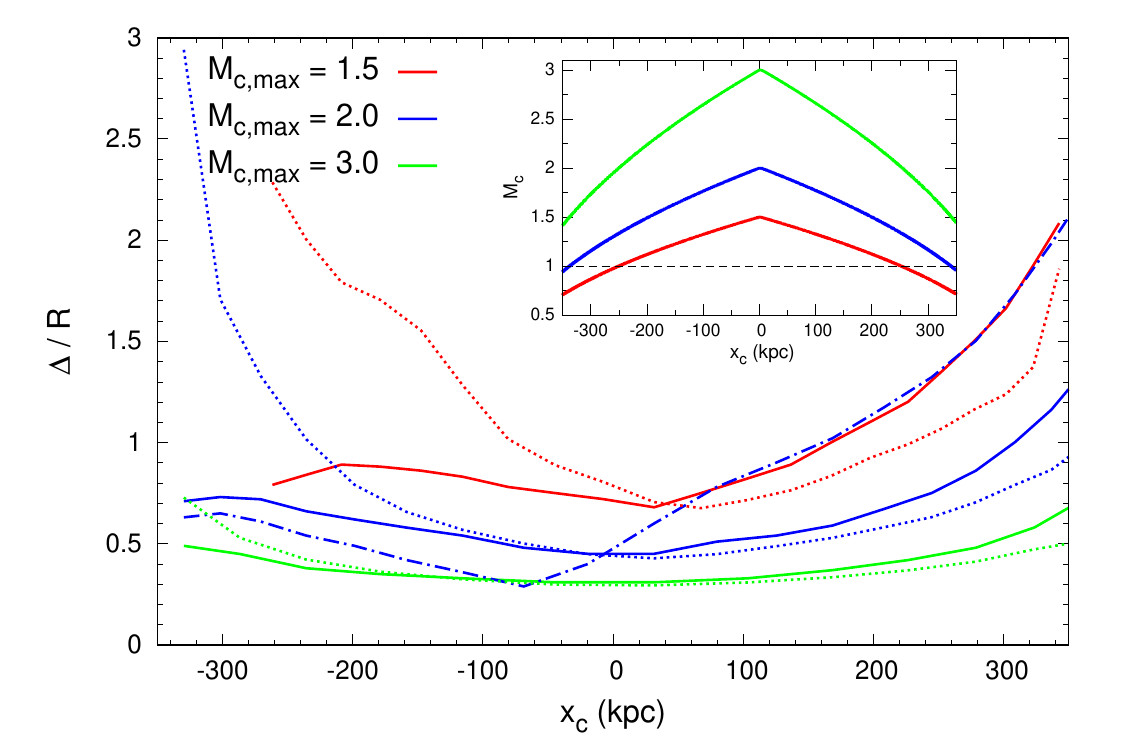}
\caption{Impact of acceleration/deceleration of the subcluster on the normalized standoff distance ($\Delta/R$) in the V-group runs. In these runs, the subcluster experiences a continuous acceleration before reaching $x=0$ and  a continuous deceleration afterwards (see equation~\ref{eq:profile_vel}). The value of $\Delta/R$ for the subcluster moving in a homogeneous medium for three different values of the acceleration/deceleration are shown with solid lines. Corresponding subcluster Mach numbers $\mathcal{M}_{c}$ are shown in the inlay (colored lines, the horizontal dashed line shows the sound speed). The maximum $\mathcal{M}_{c}$ in these runs are $\mathcal{M}_{c {\rm ,max}}=1.5,\ 2,\ 3$, respectively (see also equation~\ref{eq:profile_vel}, and Table~\ref{tab:profile_vel} for the parameters). For comparison, the dotted lines show the reference value of the standoff distance $\Delta_{V}/R$ evaluated using the shock Mach number. From this comparison, it is clear that acceleration tends to move the bow shock closer to the subcluster, while deceleration has an opposite effect (see Section~\ref{sec:rigid:gravity:acceleration}). Finally, the dashed-dotted blue line show the combined effect of the varying velocity (see the corresponding velocity evolution in the inlay -- blue curve) and of the varying density (G5 run), discussed in Section~\ref{sec:rigid:gravity:combination}. Both effects work in the same direction, causing the reduction of $\Delta/R$ before the core passage and the increase of $\Delta/R$ afterwards.
 }
\label{fig:foam_dist_uni}
\end{figure}

We now consider the case when the effects of acceleration/deceleration are combined with the increasing/decreasing density (pressure) profiles, as expected to occur in real minor mergers. The dotted-dashed line in Figure~\ref{fig:foam_dist_uni} shows the result from a run with varying velocity, but for a subcluster moving in a medium with varying density and pressure (run G5, where the time variations of velocity are taken from the V2 run, where the velocity reaches $\mathcal{M}_{c{\rm ,max}}=2$). Similar to the deviation from the baseline value seen in Figure~\ref{fig:foam_dist_icm}, $\Delta/\Delta_{V}<1$ when $x_{\rm c}<0$ and $\Delta/\Delta_{V}>1$ when $x_{\rm c}\gtrsim0$. As expected, both effects work in the same direction, since the sign of the gradients and the gravitational acceleration are closely related (at least in an isothermal atmosphere), causing underestimation/overestimation of the Mach number from the value of $\Delta/R$ assuming the reference model \citep[e.g.][]{Verigin2003}. Moreover, we see that the combination of the two effects produces an even larger deviation than the linear sum of the deviations expected from each effect separately. These strong deviations open the possibility to distinguish the pre- and post-merger phases in observations, provided that the X-ray surface brightness analysis offers an independent estimate of the Mach number, at least when the merger axis is close to the plane of sky or if the merger axis angle to the line of sight (LOS) can be inferred (see more discussions in Section~\ref{sec:rigid:projection}).

\begin{table}
\begin{minipage}{0.45\textwidth}
\centering
\caption{Parameters adopted in equation~(\ref{eq:profile_vel}).}
\label{tab:profile_vel}
\begin{tabular}{cccc}
\hline \hline
IDs & $a_{\rm x}\ (\rm kpc\Gyr^{-2})$ & $t_{\rm p}\ (\rm Gyr)$ & $\mathcal{M}_{c{\rm ,max}}$\footnote{The maximum subcluster Mach number.}\\\hline
V1 & 700.7 & 1.13 & 1.5 \\\hline
V2/G5 & 1245.6 & 0.85 & 2 \\\hline
V3 & 2802.6 & 0.57 & 3 \\
\hline\hline
\vspace*{-5mm}
\end{tabular}
\end{minipage}
\end{table}

\subsection{Effect of Subcluster Shape} \label{sec:rigid:shape}

In the previous section, we modelled the subcluster as a rigid sphere. In real merging systems, the subclusters often show a bullet-like shape (e.g. A115; \citealt{Hallman2018}, A3667; \citealt{Sarazin2016}). When measuring the normalized standoff distance $\Delta/R$, the curvature radius of the body's nose is often used as $R$ \citep[e.g.][]{Dasadia2016}. In this section, we test the validity of this approximation.

\begin{figure}
\centering
\includegraphics[width=0.45\textwidth]{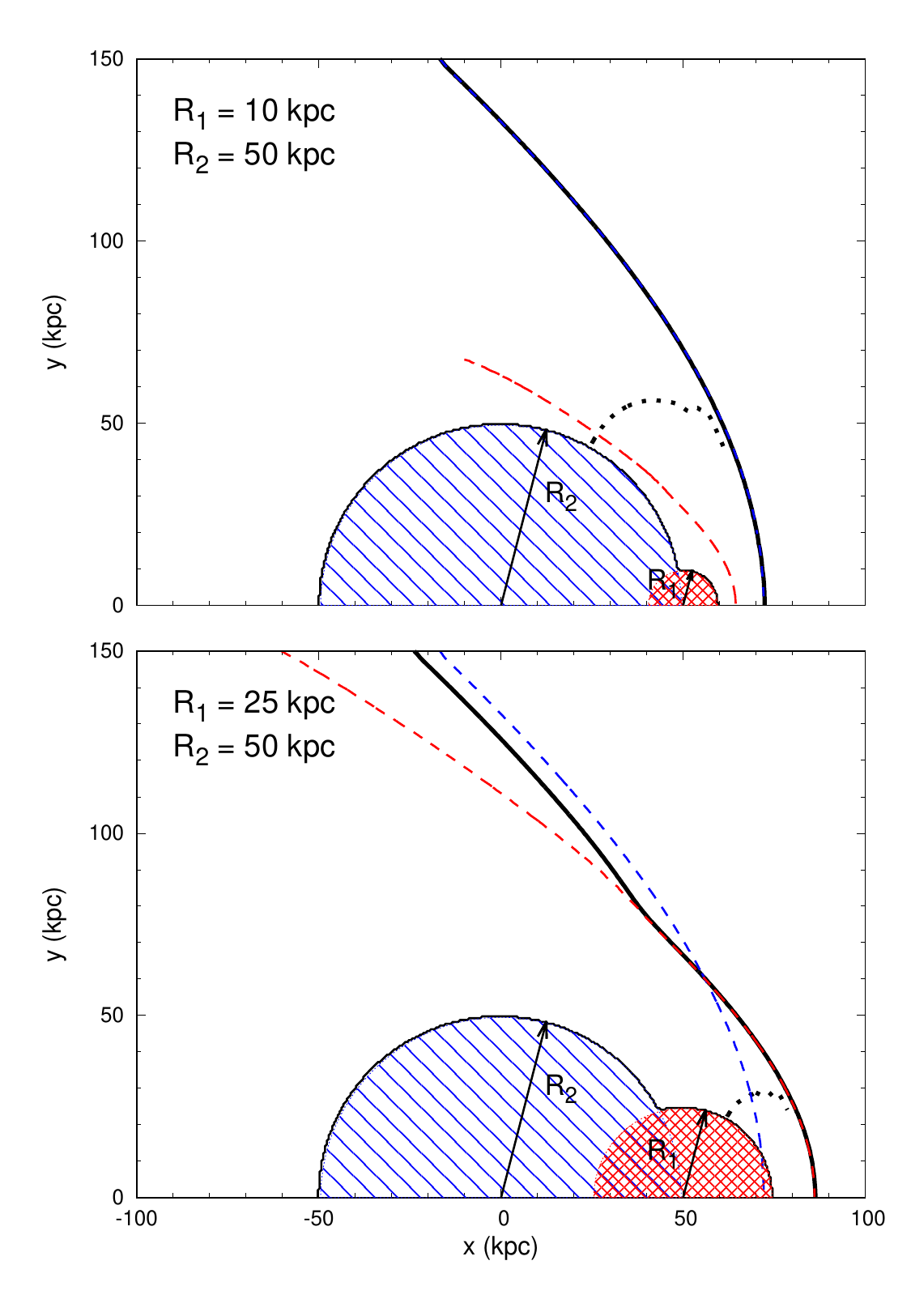}
\caption{Bow shocks driven by the rigid subcluster whose shape can be approximated as a combination of two spheres with radii $R_1$ and $R_2$ respectively (thick black lines, S-group runs). Black dotted lines show the sonic lines, where the Mach number of the shock-heated flow is unity. Red and blue dashed lines depict the bow shocks that would be driven by these spheres separately (the blue curve overlaps completely with the black one in the top panel). This figure shows that the head of the shock coincides with the bow shock produced by one of the spheres that is farther away from the tip of the subcluster (see Section~\ref{sec:rigid:shape}). }
\label{fig:foam_shape}
\end{figure}

\begin{figure}
\centering
\includegraphics[width=0.45\textwidth]{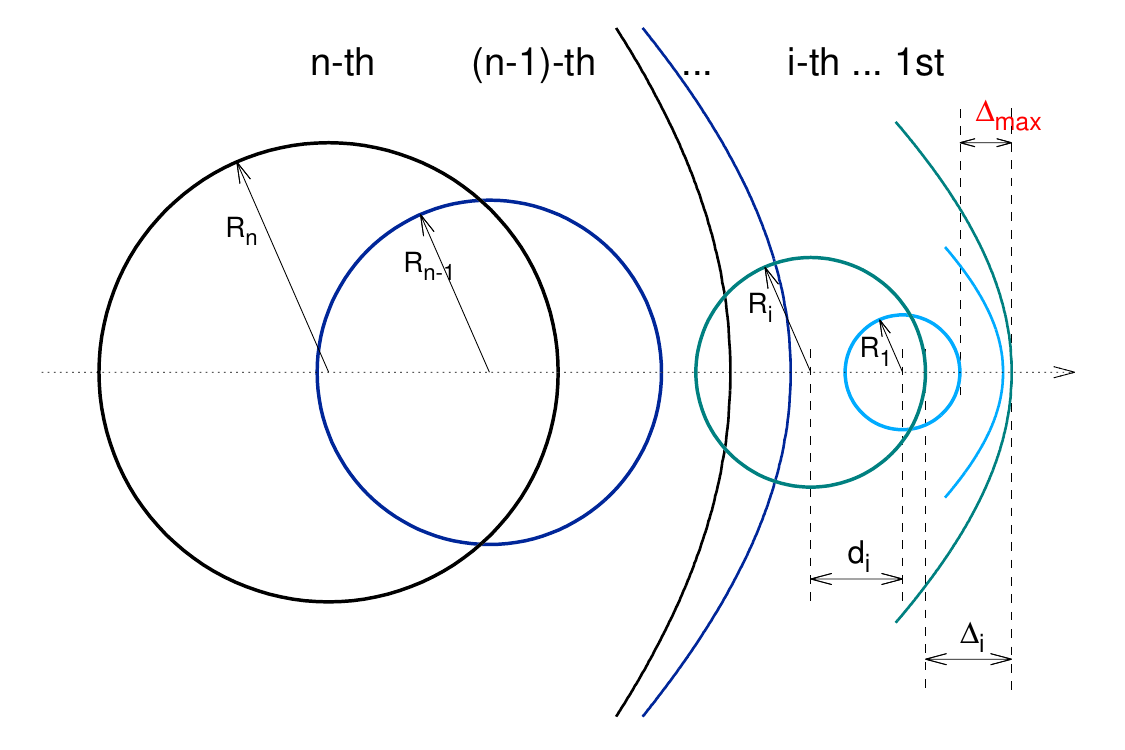}
\caption{Sketch showing a scheme of estimating the standoff distance $\Delta_{\rm max}$ for a rigid subcluster that can be decomposed into $n$ spheres with radii $R_i\ (i=1-n)$. Curves depict the bow shock fronts generated by the corresponding spheres (same color), respectively. Determination of $\Delta_{\rm max}$ is expressed in equation~(\ref{eq:delta_max}), see Section~\ref{sec:rigid:shape}.}
\label{fig:scheme_shape}
\end{figure}

For simplicity, we consider a special class of shapes that can be decomposed into a combination of two spheres. Figure~\ref{fig:foam_shape} shows two examples of the shape and the position of the bow shock and the sonic line (black thick solid lines and dotted lines) driven by a rigid subcluster moving in the uniform medium with constant velocity ($\mathcal{M}_c=2$, S-group runs). For comparison, the expected bow shocks that would be driven by the individual spheres are shown with the dashed lines. One can see that, in both panels, the leading part of the bow shock coincides with one of the dashed lines (the one that is farther away from the tip of the body) in both shape and position. Therefore, to describe the leading part of the shock, it is sufficient to compute the bow shocks for each sphere and select the leading bow shock. The sonic point (viz. intersection of the sonic line and the body surface) also lies on this corresponding sphere, generally consistent with the arguments in \citet{Moekel1949}. The radius of this corresponding sphere should be used as $R$ when calculating the normalized standoff distance $\Delta/R$ \citep{Dyke1958,Farris1994}. This procedure could be generalized to the situation when the body is represented by a combination of $n$ spheres (only for a blunt body, see Fig.~\ref{fig:scheme_shape}). Namely, for a known (assumed) shock Mach number $\mathcal{M}$, finding a maximum of the expression
\be
\Delta_{\rm max} = \max \left \{R_i\times \left [1+\frac{\Delta}{R}(\mathcal{M})\right]-R_1-d_i \right \},\ \ \ i=1-n,
\label{eq:delta_max}
\ee
where $d_i$ is the shift of the $i$-th sphere center with radius $R_i$ with respect to the the 1st sphere at the tip of the body; $\frac{\Delta}{R}(\mathcal{M})$ is the normalized standoff distance as a function of the Mach number, e.g. the expression from \citet[][]{Verigin2003}, see also equation~(\ref{eq:proxy_delta}) in Appendix~\ref{sec:appendix:proxy}. The resulting value of $\Delta_{\rm max}$ and the $R_i$ for which the maximum is achieved, can then be used to verify that the Mach number used in the above expression is consistent with the observed position of the merger shock. From Figure~\ref{fig:scheme_shape}, it is clear that using the curvature radius $R_1$ could significantly overestimate the normalized standoff distance, especially for a mildly supersonic merger.

\begin{figure}
\centering
\includegraphics[width=0.45\textwidth]{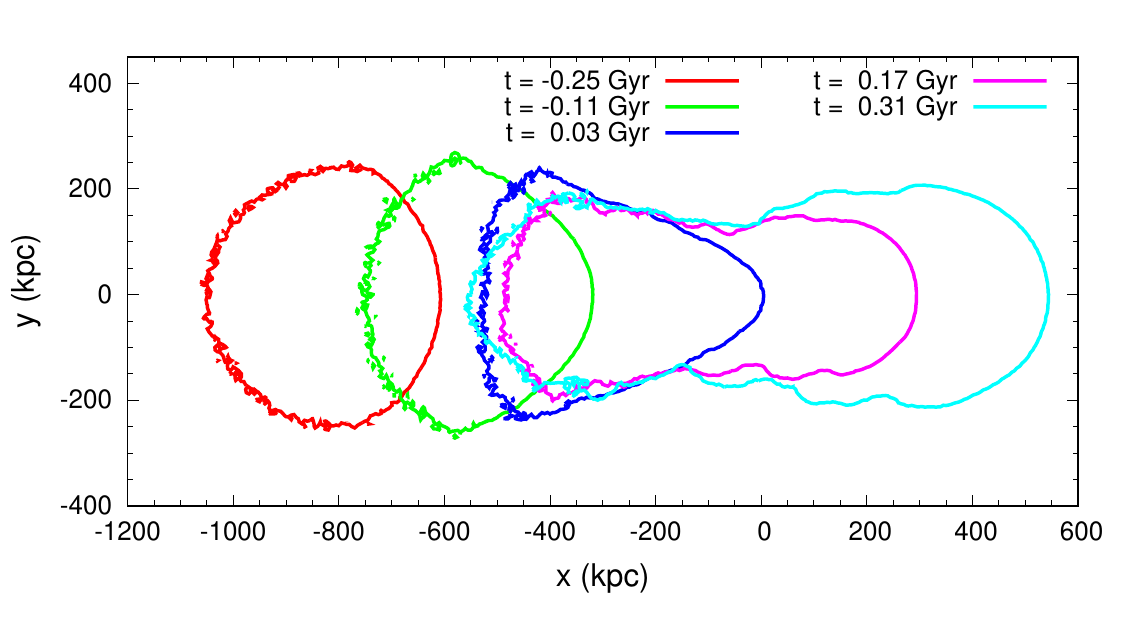}
\caption{Iso-density surfaces ($=3\times10^{-26} {\rm\,g\,cm^{-3}}$) of gas of the infalling subcluster in the merger plane at different times (obtained from the SPH simulation shown in Fig.~\ref{fig:sph_slice}). This figure shows the evolution of the subcluster shape during the merger process (see Section~\ref{sec:rigid:shape}).}
\label{fig:sph_shape}
\end{figure}

Finally, we discuss the deformation of the merging subcluster and its effect on the standoff distance, which could not be captured by our rigid-subcluster model. Figure~\ref{fig:sph_shape} shows the evolution of the infalling subcluster shape (i.e. isodensity surfaces in the merger plane) in our SPH simulation (cf. the mesh-based simulation results presented in \citealt{Roediger2015}). During the merger process, the subcluster shape varies with the time. When the subcluster approaches the pericenter, it is elongated and has a bullet-like shape. After core passage, the head of the subcluster is generally round but expands when it moves outwards because the outside pressure is decreasing. This result, of course, depends on the merger parameters used in the simulation (e.g. mass ratio, impact parameter, and relative velocity). We could roughly estimate the rate of change of the curvature radius of the head of the subcluster from Figure~\ref{fig:sph_shape}, i.e. $\sim500\kms$, much slower than the local speed of sound.

\subsection{Projection Effects} \label{sec:rigid:projection}

The effect of projection is one of the major observational issues that hinders accurate determination of the shock parameters. If the merger axis is not in the sky plane, the LOS tangent to the bow shock surface will not go through the stagnation point, leading to biases in the measured standoff distance, as well as in the derived Mach number from the density and/or temperature jumps (from the X-ray observations).

\begin{figure}
\centering
\includegraphics[width=0.45\textwidth]{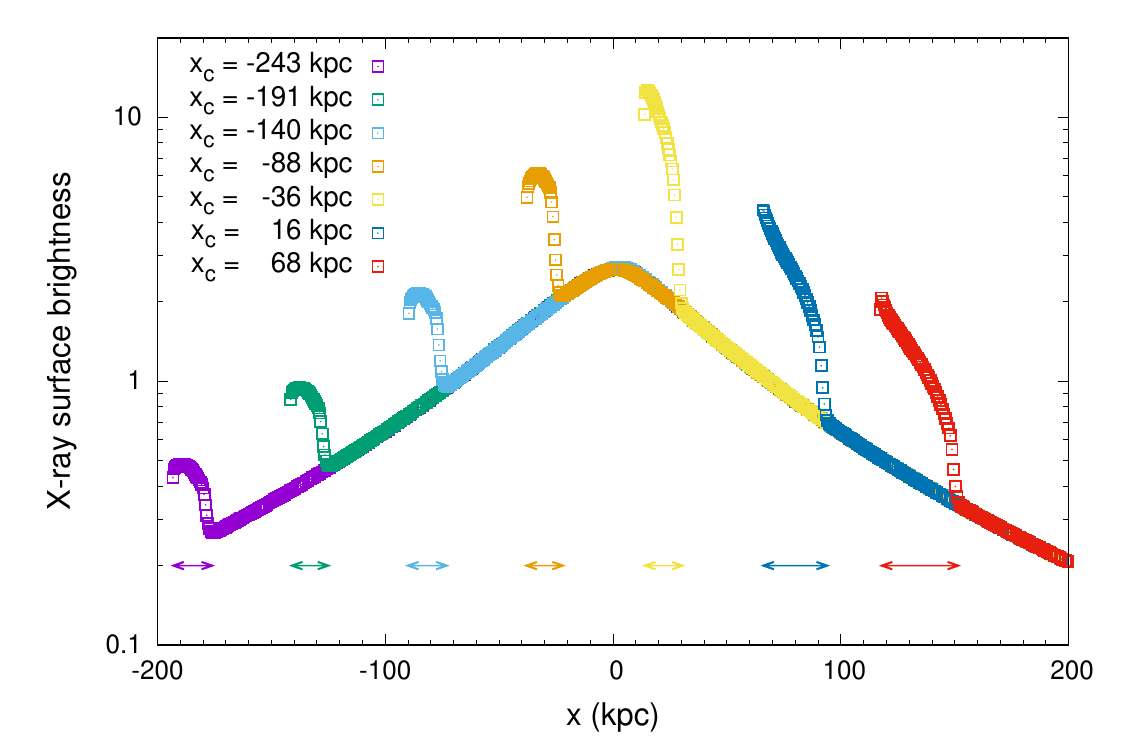}
\caption{Evolution of the profiles of the X-ray surface brightness crossing the shock front in the simulation G2 ($\theta=90^\circ$; again, $x_{\rm c}$ represents the horizontal position of the subcluster center at the given moment). Their standoff distances are marked by the arrows in the corresponding color. This figure shows the surface brightness `jump' associated with the shock is less prominent in the infall phase and slightly more complicated to model (see Section~\ref{sec:rigid:projection}).}
\label{fig:foam_xray}
\end{figure}

In this section, we model the profile of the X-ray surface brightness from the simulations along a given LOS\footnote{The inclination angle $\theta$ is defined as the angle between the LOS and the merger axis, e.g. $\theta=\pi/2$ corresponds to the merger in the sky plane.} as
\be
S_{\rm X}(r_{\rm p})\propto\int_{\rm LOS}{\rho_{\rm gas}^2{\rm d}\ell},
\label{eq:xray}
\ee
where $\rho_{\rm gas}$ is the gas density; $r_{\rm p}$ is the projected distance from the subcluster center along the merger axis. No dependence of the gas emissivity on the temperature is included in this equation, because we  focus on the soft X-ray band. As an example, the evolution of the X-ray surface brightness profiles in the simulation G2 (for $\theta=90^\circ$, i.e. the merger is in the sky plane) is shown in Figure~\ref{fig:foam_xray}. During the infall phase, the surface brightness of the unperturbed main cluster increases ahead of the infalling subcluster ($x<0$ in Fig.~\ref{fig:foam_xray}). This leads to a lower apparent amplitude of the surface brightness `jump' associated with the shock and, in addition, slightly more complicated modelling. We therefore focus only on the phase after core passage ($x>0$ in Fig.~\ref{fig:foam_xray}).

\begin{figure}
\centering
\includegraphics[width=0.45\textwidth]{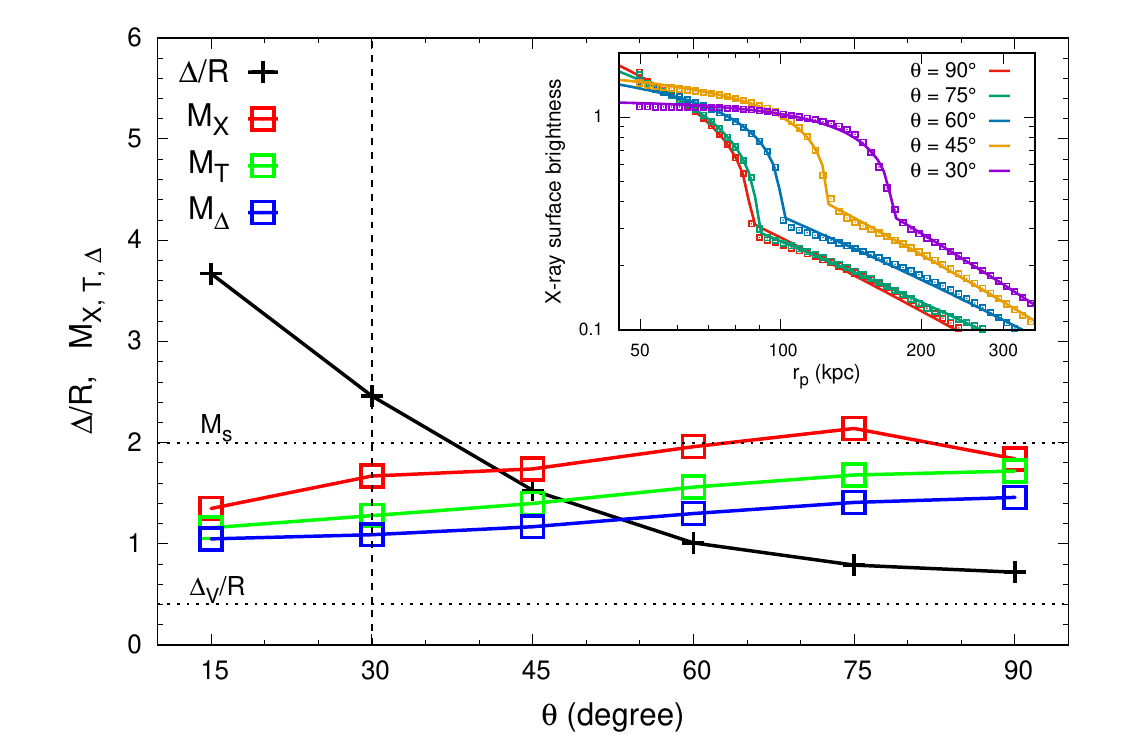}
\caption{Normalized standoff distance and Mach numbers (determined from the X-ray surface brightness, X-ray weighted temperature profiles, and the standoff distance, respectively) as a function of the inclination angle $\theta$ in the simulation G2, when the subcluster reaches $x_{\rm c}=224\kpc$. The horizontal black dotted lines mark the shock Mach number $\mathcal{M}_s(=2)$ and the corresponding $\Delta_{V}/R$. The vertical black dashed line marks the corresponding Mach angle. The profiles of the X-ray surface brightness (points) and the best-fit models (solid lines) are shown in the inset. This figure shows the significant projection effect on the standoff distance, which is a decreasing function of the inclination angle (see Section~\ref{sec:rigid:projection}).}
\label{fig:foam_mach_angle}
\end{figure}

\begin{figure}
\centering
\includegraphics[width=0.45\textwidth]{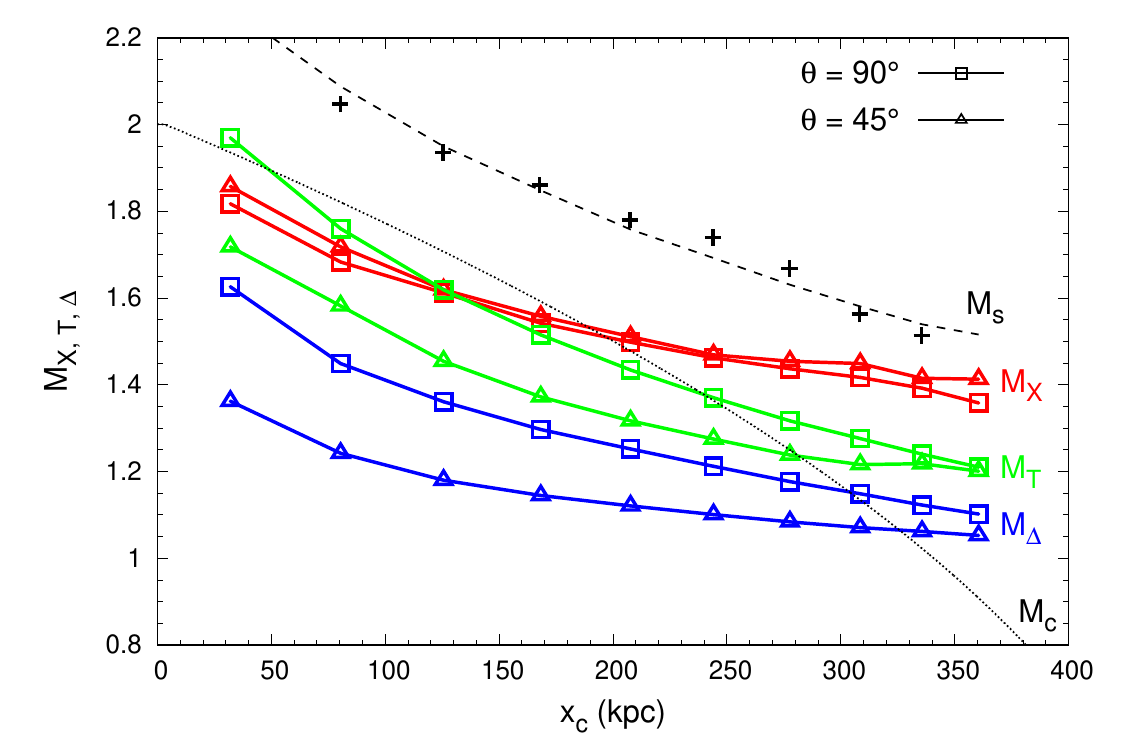}
\caption{Time evolution of the Mach numbers $\mathcal{M}_{X},\ \mathcal{M}_{T}$ and $\mathcal{M}_{\Delta}$ estimated from the X-ray surface brightness, X-ray weighted temperature profiles, and the normalized standoff distance, respectively, at $\theta=90^\circ$ and $45^\circ$, in the simulation G5. As a comparison, the dotted and dashed lines show the evolution of Mach numbers of the subcluster $\mathcal{M}_c(=\mathcal{M}_g)$ and the shock front $\mathcal{M}_s$, respectively; the black pluses show the shock Mach number measured directly from the jump of gas temperature across the shock (but not from the X-ray weighted temperature). This figure shows the strong impact on measuring $\mathcal{M}_s$ in the post-merger stage, caused by projection effects (see Section~\ref{sec:rigid:projection}).}
\label{fig:foam_mach_time}
\end{figure}

We explore the dependence of the X-ray profile on $\theta$. Figure~\ref{fig:foam_mach_angle} shows the results of the G2 run, when the rigid subcluster reaches $x_{\rm c}=224\kpc$. A broken power-law model is used to fit the X-ray profiles \citep[e.g.][]{Markevitch2007}, which provides the best-fit standoff distance and the X-ray Mach number $\mathcal{M}_{X}$ (from the shock compression). As expected, given the 3D shape of the bow shock, the standoff distance is a decreasing function of the inclination angle $\theta$, which is particularly sensitive to $\theta$ when $\theta<60^\circ$. Even when $\theta$ is smaller than the Mach angle, it is still possible to observe the shock front in the X-ray profile. That is because the edge of the Mach cone is not straight in the simulations, but bends towards the wake of the subcluster. 
Comparison of the Mach numbers $\mathcal{M}_{X},\ \mathcal{M}_{T}$, and $\mathcal{M}_{\Delta}$ determined from the X-ray surface brightness, X-ray weighted temperature\footnote{The X-ray weighted temperature is defined as $\int_{\rm LOS}{T_{\rm gas}\rho_{\rm gas}^2{\rm d}\ell}/\int_{\rm LOS}{\rho_{\rm gas}^2{\rm d}\ell}$, where $T_{\rm gas}$ is the gas temperature. We simply use the ratio between the X-ray weighted temperature on the downstream and upstream sides of the shock to estimate $\mathcal{M}_T$ (see equation~\ref{eq:proxy_temp}).} profiles, and the normalized standoff distance, respectively, at different viewing angles, is shown in Figure~\ref{fig:foam_mach_angle}. One can see that, $\mathcal{M}_{\Delta}$ clearly underestimates the Mach number of the shock. Compared to $\mathcal{M}_{T}$ and $\mathcal{M}_{\Delta}$, $\mathcal{M}_{X}$ shows a relatively weak dependence on the inclination angle (see also Fig.~\ref{fig:foam_mach_time}) and provides a robust estimate of the Mach number if $\theta$ is not too small ($\theta\gtrsim60^\circ$).

Figure~\ref{fig:foam_mach_time} shows the time evolution of the Mach numbers $\mathcal{M}_{X},\ \mathcal{M}_{T}$ and $\mathcal{M}_{\Delta}$ estimated at $\theta=90^\circ$ and $45^\circ$ in the simulation G5. These results are similar with those in Figure~\ref{fig:foam_mach_angle}. The standoff distance gives the smallest Mach number. $\mathcal{M}_{X}$ still shows the weakest dependence on the inclination angle. However, all three measuring methods show more significant underestimation on the shock Mach number, because the deceleration of the subcluster makes the gas distribution behind the shock more complicated (see the prominent difference between $\mathcal{M}_s$ and $\mathcal{M}_c$ in Fig.~\ref{fig:foam_mach_time}). For comparison, the Mach number directly measured from the gas temperature discontinuity across the shock (but not from the X-ray weighted temperature) is shown as the black pluses, which is consistent with $\mathcal{M}_s$ as expected. 

\section{Conclusion} \label{sec:conclusion}

In this paper, we have explored various effects that influence the position of a bow shock relative to the contact discontinuity (a.k.a. cold front) that drives this shock in a context of a cluster merger, specialized for the case of a minor merger. Unlike the textbook case of a bow shock ahead of a small body moving in a homogeneous gas, the standoff distance is not only a function of the body size and velocity, but it is also sensitive to the cluster environment. Therefore, the standoff distance is not only as a complementary proxy for the merger Mach number (see Section~\ref{sec:introduction} for the discussion of different Mach numbers present in the problem), but it also provides information on the merger configuration, that is difficult to obtain from, e.g. the Mach number of the bow shock. Our findings can be summarized as follows.
\begin{itemize}
\item The standoff distance $\Delta$ is a simple geometrical proxy for the velocity of a supersonically moving body in a uniform medium, which we refer to as a `standard' case. It is a sensitive function of the body Mach number, as long as it is smaller than $\sim 3$. In the context of merging galaxy clusters, it provides a useful and non-trivial diagnostic of the merger process, but its relation with the velocity of the body is complicated.

\item One important difference with the standard case is that the velocity of a subcluster does not necessarily coincide with the velocity of the shock it drives (see Sections~\ref{sec:sph} and \ref{sec:rigid:gravity}). This is caused by two effects, namely, (i) the acceleration (deceleration) of the subcluster by the gravitational pull of the main cluster before (after) the core passage and (ii) motion in (against) the direction of density and pressure gradients before (after) the core passage. These two effects work in the same direction and tend to push the shock closer to (farther away from) the body before (after) the core passage. The magnitude of these effects scales with the size of the gaseous core of the subcluster; for a subcluster that is much smaller than the characteristic scale height of the main cluster atmosphere, the effects are small.
    This means that, before the core passage, the shock Mach number is fairly close to the subcluster (gas) Mach number; while after core passage, the shock can be moving faster (and be farther away) than the subcluster. Therefore, if the analysis of the density/temperature jumps at the shock yield a Mach number significantly larger than the one inferred from the standoff distance, one can then conclude that the subcluster has already passed through the densest part of the main cluster. These effects provide an explanation for an unexpectedly large standoff distance seen in some observations \citep[e.g.][]{Dasadia2016}.

\item  The non-trivial shape of the infalling subcluster causes ambiguities in determining the normalized standoff distance (see Section~\ref{sec:rigid:shape}). We propose a method to address this issue by decomposing the geometry of the cluster into a series of spheres. The location of the merger shock could be determined by comparing the bow shocks for each sphere and selecting the leading one (see the scheme in Fig.~\ref{fig:scheme_shape}).

\item We analyzed the impact of the projection effects on the Mach number proxies, specializing for a subcluster after core passage (see Section~\ref{sec:rigid:projection}). Projection effects tend to underestimate the shock Mach number derived from the density and temperature jumps unless the merger is exactly in the sky plane. The Mach number derived from the surface brightness analysis appears to be the least affected (see Fig.~\ref{fig:foam_mach_time}). The standoff distance analysis produces the smallest Mach number, partly because of the effects mentioned above.

\end{itemize}

Finally, we emphasize that this study is not intended as a replacement for full dark matter plus gas simulations, that automatically include all the effects discussed above. Instead, this study could be used to guide the interpretation of the observational data, provide initial guess on the merger configuration, and help setting up full hydrodynamic simulations.

\section*{Acknowledgements}

EC acknowledges support by grant No.~14-22-00271 from the Russian Scientific Foundation. WF and CJ acknowledge support from the Smithsonian Institution, NASA Contract NAS-08060, and NASA Grant GO2-13005X.

\appendix

\section{Observational proxies for Mach number} \label{sec:appendix:proxy}

For convenience of the reader, in this section, we summarize the RH relations for the gas properties (density, temperature, and pressure)\footnote{Throughout this paper, the gas is assumed to be ideal monatomic with adiabatic index $\gamma=5/3$.} on the downstream and upstream sides of the shock \citep{Landau1959}, i.e.
\be
\frac{\rho_{\rm d}}{\rho_{\rm u}} = \frac{4\mathcal{M}_s^2}{\mathcal{M}_s^2+3},
\ee
\be
\frac{T_{\rm d}}{T_{\rm u}} = \frac{5\mathcal{M}_s^4+14\mathcal{M}_s^2-3}{16\mathcal{M}_s^2},
\label{eq:proxy_temp}
\ee
\be
\frac{P_{\rm d}}{P_{\rm u}} = \frac{5\mathcal{M}_s^2-1}{4},
\ee
and also the baseline model of the normalized standoff distance for a rigid sphere moving in the homogenous medium (i.e. equation~35 in \citealt{Verigin2003} with $b_0=-1$),
\be
\frac{\Delta_V}{R}=A\epsilon^{2/3}\Big(1-\frac{B}{\epsilon^{1/6}}\Big),
\label{eq:proxy_delta}
\ee
where $A=0.956$, $B=0.473$, and $\epsilon=(\mathcal{M}_s^2+3.213)/(3\mathcal{M}_s^2-3)$.
Figure~\ref{fig:mach_proxy} shows these relations for $\mathcal{M}_s$ in the range from $1$ to $5$.

\section{SPH simulations} \label{sec:appendix:sph}

We performed SPH simulations of a merger between two idealized galaxy clusters by using the Gadget-2 code \citep{Springel2001}. The methodology has been described in \citet{Zhang2014,Zhang2015} in detail.
Here, we give a brief summary.
\begin{itemize}
  \item In the simulations, each of the two merging clusters consists of a spherical DM halo and gas halo (see equations~2 and 3 in \citealt{Zhang2014} for their density profiles). The baryonic mass fraction within the virial radius is fixed to be $0.15$. In the initial conditions, each cluster is modelled in equilibrium. We have tested that a single cluster could maintain its initial stable configuration in the simulations for more than $10\Gyr$.
  \item Cartesian coordinates $(x,\ y,\ z)$ are adopted for our SPH simulations. The center of mass of the merging system is initially at rest at the origin of the coordinate system. The merger plane coincides with the $x-y$ plane. The initial distance between two merging clusters is set to twice the sum of their virial radii. Four parameters are defined to describe the merger configuration, including the masses of the main cluster $M_{\rm 1st}$ and the subcluster $M_{\rm 2nd}$ ($M_{\rm 1st}>M_{\rm 2nd}$), the initial relative velocity ($V_0$), and the impact parameter ($P_0$). These of the minor merger presented in Section~\ref{sec:sph} are summarized in Table~\ref{tab:sph_params}.
\end{itemize}

\begin{table}
\begin{center}
\caption{Parameters of SPH simulations.}
\label{tab:sph_params}
\begin{tabular}{cccc}
\hline \hline
$M_{\rm 1st}\ (\msun)$ & $M_{\rm 2nd}\ (\msun)$ & $V_0\ (\kms)$ & $P_0\ (\kpc)$ \\\hline
$10^{15}$ & $10^{14}$ & $500$ & $0$ \\
\hline\hline
\end{tabular}
\end{center}
\end{table}

\section{Simulations of rigid subclusters} \label{sec:appendix:openfoam}

We carry out our rigid-subcluster simulations by using the mesh-based code OpenFOAM\footnote{Open Source Field Operation and Manipulation, www.openfoam.org.}. The built-in sonicDyMFoam solver is selected to handle the hydrodynamic equations. 

The simulations are done in a two-dimensional (2D) axisymmetric coordinate system $(x,\ y)$. The axis of symmetry is along the $x$-axis (also assumed as the merger axis). For all our simulations, the computational domain is set to be $x\in[-600,\ 600\kpc]$ and $y\in[0,\ 600\kpc]$, sufficiently large to resolve the shock Mach cone. The rigid subcluster is modelled as a wall with a slip boundary condition. The effective spatial resolution reaches $0.75\kpc$ within the two-radius region around the subcluster. A reflecting boundary condition is adapted at the edges of the domain in both of the $x$ and $y$ directions. The k-omega-SST Reynolds-averaged simulation (RAS) modelling is employed to handle the subgrid turbulence.

We have performed three groups of simulations, i.e. V-, G-, and S-groups, to investigate the various causes that affect the standoff distance in galaxy clusters (see Table~\ref{tab:foam_params}). In V- and G-group runs, the subcluster is modelled as a sphere, whose radius and initial position of the center are fixed as $R=50\kpc$ and $(-450,\ 0\kpc)$. In S-group runs, a non-spherical shape of the rigid subclusters is implemented (see Section~\ref{sec:rigid:shape}).

In G-group runs, a fixed gravitational potential is assumed to model the cluster environment,
\be
  \Phi(r) = 2v_{\rm c}^{2}\ln(r+r_{\rm c}),
\label{eq:profile_pot}
\ee
where $r=\sqrt{x^2+y^2}$, $v_{\rm c}=400\kms$ and $r_{\rm c}=100\kpc$ are the scaling parameter for potential and the core radius, respectively. The ICM is initially assumed to be isothermal and in hydrostatic equilibrium. The corresponding gas density is
\be
  \rho_{\rm gas}(r) = \rho_{\rm c}\exp\Big[-\frac{\Phi(r)}{{c_{T}}^2}\Big],
\label{eq:profile_rho}
\ee
where $\rho_{\rm c}=6.77\times10^{-22}{\,\rm g\,cm^{-3}}$ and ${c_{T}}\equiv\sqrt{k_{\rm B}T_{\rm gas}/\mu m_{\rm p}}$ are the scale density and the isothermal sound speed of the gas, respectively; $T_{\rm gas}=1\keV$, $k_{\rm B}$, $\mu\,(=0.6)$, and $m_{\rm p}$ are the initial gas temperature, Boltzmann constant, mean molecular weight per ion, and proton mass, respectively. The solver is modified to include this static gravitational field. It is worth noting that the above parameters define the environment a cool galaxy group, however, in terms of the goals of this study, these settings would have little effect on our final conclusions (e.g. for the more massive clusters).

\begin{table*}
\centering
\begin{minipage}{0.9\textwidth}
\centering
\caption{Parameters of rigid-subcluster simulations.}
\label{tab:foam_params}
\begin{tabular}{cccccccc}
  \hline\hline
 IDs &
 Atmosphere\footnote{The environment where the subcluster moves.}&
 Shape\footnote{The shape of the infalling rigid subcluster.} &
 Radius/radii ($\rm kpc$)\footnote{The subcluster radius $(R)$, or radii $(R_1/R_2)$ if the subcluster's shape can be approximated as a combination of two spheres.} &
 Type\footnote{The motion type of the rigid subclusters: moving at time-dependent velocity (the `varying' type; see equation~\ref{eq:profile_vel} and Table~\ref{tab:profile_vel}) or at constant velocity (the `constant' type). }&
 $\mathcal{M}_c$\footnote{The Mach number of the rigid subcluster moving in `constant' type. } &\\
\hline
\multicolumn{6}{c}{Spherical subclusters moving in a uniform atmosphere} \\
\multicolumn{6}{c}{with varying velocity (V-group)} \\
\hline
  V1/V2/V3 & uniform & sphere & 50 & varying & - &  \\
\hline
\multicolumn{6}{c}{Spherical subclusters moving in a stratified atmosphere} \\
\multicolumn{6}{c}{with density (pressure) gradients (G-group)}\\
\hline
  G1 & ICM &  sphere & 50 & constant & 1.5 & \\\hline
  G2 & ICM &  sphere & 50 & constant & 2 & \\\hline
  G3 & ICM &  sphere & 50 & constant & 3 & \\\hline
  G4 & ICM &  sphere & 12.5 & constant & 2 & \\\hline
  G5 & ICM &  sphere & 50 & varying  & - & \\
\hline
\multicolumn{6}{c}{Non-spherical subclusters moving in a uniform atmosphere}\\
\multicolumn{6}{c}{with constant velocity (S-group)}\\
\hline
  S1 & uniform & peanut & $10/50$ & constant & 2 & \\\hline
  S2 & uniform & peanut & $25/50$ & constant & 2 & \\
\hline\hline
\vspace*{-5mm}
\end{tabular}
\end{minipage}
\end{table*}


\bsp	
\label{lastpage}

\begin{thebibliography}{0}
\expandafter\ifx\csname natexlab\endcsname\relax\def\natexlab#1{#1}\fi

\bibitem[Botteon et al.(2018)]{Botteon2018} Botteon, A., Gastaldello, F., \& Brunetti, G.\ 2018, \mnras, 476, 5591

\bibitem[Bykov et al.(2015)]{Bykov2015} Bykov, A.~M., Churazov, E.~M., Ferrari, C., et al.\ 2015, \ssr, 188, 141

\bibitem[Dasadia et al.(2016)]{Dasadia2016} Dasadia, S., Sun, M., Morandi, A., et al.\ 2016, \mnras, 458, 681

\bibitem[Fairfield et al.(2001)]{Fairfield2001} Fairfield, D.~H., Cairns, I.~H., Desch, M.~D., et al.\ 2001, \jgr, 106, 25361

\bibitem[Farris \& Russell(1994)]{Farris1994} Farris, M.~H., \& Russell, C.~T.\ 1994, \jgr, 99, 17

\bibitem[Hallman et al.(2018)]{Hallman2018} Hallman, E.~J., Alden, B., Rapetti, D., Datta, A., \& Burns, J.~O.\ 2018, arXiv:1804.06493
    
\bibitem[Keshet \& Naor(2016)]{Keshet2016} Keshet, U., \& Naor, Y.\ 2016, \apj, 830, 147

\bibitem[Landau \& Lifshitz(1959)]{Landau1959} Landau, L.~D., \& Lifshitz, E.~M.\ 1959, Course of theoretical physics, Oxford: Pergamon Press, 1959

\bibitem[Markevitch et al.(2002)]{Markevitch2002} Markevitch, M., Gonzalez, A.~H., David, L., et al.\ 2002, \apjl, 567, L27

\bibitem[Markevitch et al.(2005)]{Markevitch2005} Markevitch, M., Govoni, F., Brunetti, G., \& Jerius, D.\ 2005, \apj, 627, 733

\bibitem[Markevitch \& Vikhlinin(2007)]{Markevitch2007} Markevitch, M., \& Vikhlinin, A.\ 2007, \physrep, 443, 1

\bibitem[Moekel(1949)]{Moekel1949} Moekel. W. E. 1949, Approximate Method for Predicting Form and Location of Detached Shock Waves Ahead of Plane or Axially Symmetric Bodies, NACA Technical Note 1921

\bibitem[Petrinec(2002)]{Petrinec2002} Petrinec, S.~M.\ 2002, \planss, 50, 541

\bibitem[Roediger et al.(2015)]{Roediger2015} Roediger, E., Kraft, R.~P., Nulsen, P.~E.~J., et al.\ 2015, \apj, 806, 103

\bibitem[Sarazin et al.(2016)]{Sarazin2016} Sarazin, C.~L., Finoguenov, A., Wik, D.~R., \& Clarke, T.~E.\ 2016, arXiv:1606.07433

\bibitem[Springel et al.(2001)]{Springel2001} Springel, V., Yoshida, N., \& White, S.~D.~M.\ 2001, \na, 6, 79

\bibitem[Springel \& Farrar(2007)]{Springel2007} Springel, V., \& Farrar, G.~R.\ 2007, \mnras, 380, 911

\bibitem[Russell et al.(2010)]{Russell2010} Russell, H.~R., Sanders, J.~S., Fabian, A.~C., et al.\ 2010, \mnras, 406, 1721

\bibitem[Van Dyke(1958)]{Dyke1958} Van Dyke, M. D.\ 1958, Journal of the Aerospace Sciences, 25, 485

\bibitem[Verigin et al.(2003)]{Verigin2003} Verigin, M., Slavin, J., Szabo, A., et al.\ 2003, Journal of Geophysical Research (Space Physics), 108, 1323

\bibitem[Vikhlinin et al.(2001)]{Vikhlinin2001} Vikhlinin, A., Markevitch, M., \& Murray, S.~S.\ 2001, \apj, 551, 160

\bibitem[We{\.z}gowiec et al.(2011)]{Wezgowiec2011} We{\.z}gowiec, M., Vollmer, B., Ehle, M., et al.\ 2011, \aap, 531, A44

\bibitem[Zhang et al.(2014)]{Zhang2014} Zhang, C., Yu, Q., \& Lu, Y.\ 2014, \apj, 796, 138

\bibitem[Zhang et al.(2015)]{Zhang2015} Zhang, C., Yu, Q., \& Lu, Y.\ 2015, \apj, 813, 129

\bibitem[Zuhone \& Roediger(2016)]{Zuhone2016} Zuhone, J.~A., \& Roediger, E.\ 2016, Journal of Plasma Physics, 82, 535820301

\end{thebibliography}
\end{document}